\documentclass[letterpaper, 10 pt, conference]{ieeeconf}  

\IEEEoverridecommandlockouts                              

\usepackage[T1]{fontenc}
\usepackage{cite}
\usepackage{amsmath,amssymb,amsfonts}
\usepackage{algorithmic}
\usepackage{graphicx}
\usepackage{textcomp}
\usepackage{xcolor}
\usepackage{comment}
\usepackage{soul}
\usepackage{amsmath}
\usepackage{booktabs}
\usepackage{lipsum}  
\usepackage{algorithm}
\usepackage{titlesec}
\usepackage{indentfirst}
\allowdisplaybreaks
\titlespacing*{\subsection}{0pt}{0.2\baselineskip}{0.2\baselineskip}
\titlespacing*{\section}{0pt}{0.2\baselineskip}{0.2\baselineskip}
\usepackage[caption=false]{subfig}

\DeclareMathOperator*{\argmin}{argmin}
\def\BibTeX{{\rm B\kern-.05em{\sc i\kern-.025em b}\kern-.08em
    T\kern-.1667em\lower.7ex\hbox{E}\kern-.125emX}}
\begin{document}
\setlength{\abovedisplayskip}{6pt}
\setlength{\belowdisplayskip}{6pt}
\title{\LARGE \bf Deep-MPC: A DAGGER-Driven Imitation Learning Strategy for Optimal Constrained Battery Charging }

\author{Jorge Espin, Dong Zhang, Daniele Toti, and Andrea Pozzi
\thanks{J. Espin and D. Zhang are with the School
of Aerospace and Mechanical Engineering, The University of Oklahoma, Norman, OK 73019, USA.  D. Zhang is also with the Institute for Resilient Environmental \& Energy Systems, The University of Oklahoma, Norman, OK 73019, USA. E-mail: {\tt\{jorge.espin, dzhang\}@ou.edu}.}
\thanks{D. Toti, and A. Pozzi are with the Faculty of Mathematics, Physical and Natural Science, Catholic University of Sacred Heart, Brescia 25133, Italy. E-mail: {\tt\{daniele.toti, andrea.pozzi\}@unicatt.it}.}
}

\maketitle

\begin{abstract}
In the realm of battery charging, several complex aspects demand meticulous attention, including thermal management, capacity degradation, and the need for rapid charging while maintaining safety and battery lifespan. By employing the imitation learning paradigm, this manuscript introduces an innovative solution to confront the inherent challenges often associated with conventional predictive control strategies for constrained battery charging. A significant contribution of this study lies in the  adaptation of the Dataset Aggregation (DAGGER) algorithm to address scenarios where battery parameters are uncertain, and internal states are unobservable. Results drawn from a practical battery simulator that incorporates an electrochemical model highlight substantial improvements in battery charging performance, particularly in meeting all safety constraints and outperforming traditional strategies in computational processing.
\end{abstract}


\section{Introduction}

The ecological transition has sparked a growing focus on batteries, particularly within the realm of sustainable mobility driven by electric vehicles (EVs)~\cite{Cairns2010batteries}. This transition underscores the vital role of batteries in promoting eco-friendly transportation. However, it also highlights the pressing need to enhance battery efficiency, long-lasting battery performance, and safety, particularly during the charging phase. To address these challenges, advanced battery management systems, often employing Model Predictive Control (MPC), have gained prominence~\cite{Torchio2015, Pozzi2018b}.

It is crucial to recognize the limitations of implementing MPC, despite its benefits. A key hurdle is its computational complexity involved in solving constrained optimization problems at each time step, especially when dealing with nonlinear battery models. Furthermore, MPC assumes perfect knowledge of system dynamics, a condition that often falls short in real-world scenarios due to uncertainties in system parameters and limitations in measuring specific internal battery states~\cite{Lu2013}.

To tackle high computational demands, researchers have resorted to utilizing reduced-order models and explicit MPC strategies. The employment of reduced-order models serves a dual purpose: it strikes a balance between model accuracy and computational complexity~\cite{ galuppini2023efficient}. Meanwhile, the adoption of explicit MPC streamlines real-time operations into fundamental function evaluations~\cite{Alessio2009_explicitMPC}. However, it is important to note that explicit MPC does exhibit limitations, particularly when dealing with an increase number of constraints~\cite{Lucia2020}. In addition to these approaches, Deep Model Predictive Control (Deep-MPC) has emerged as a promising alternative, leveraging deep neural networks to approximate predictive control laws and reduce computational burdens~\cite{Pozzi2022_ccta, Pozzi2023cace, Pozzi2023sensors}. Yet, all these methods still grapple with the difficulties of handling uncertain parameters and unobservable states.

In response to these complexities, behavioral cloning, a subtype of imitation learning in which a model mimics an expert's behavior without knowledge of the system dynamics, has been employed but struggles with distributional shift — a critical challenge in machine learning. Distributional shift occurs when the learning model encounters states that were inadequately represented or entirely absent in its training data. This challenge becomes apparent when the model deviates from the expert's path and continues to make errors that lead it into unfamiliar states, thus exacerbating the initial mistake~\cite{ross2010efficient}.  Dataset Aggregation (DAGGER) was introduced by~\cite{Ross2011dagger} as a method to address the challenge of distributional shift. This iterative algorithm aims to minimize the compounding of errors resulting from the shift by iteratively integrating the decisions made by both the learning model and an expert policy. This integration prevents the model from straying into unfamiliar areas of the state space. Through continuous inclusion of expert feedback throughout the training phase, DAGGER maintains the learning model's trajectory in close alignment with that of the expert, effectively diminishing the error attributed to distributional shifts. The application and adaptation of DAGGER to the battery charging domain, especially in scenarios involving uncertain parameters and unobservable states, represent an innovative step and constitute a significant contribution to this research.

This study proposes a novel DAGGER-based strategy for optimal battery charging, addressing distributional shifts, reducing computational complexity through offline processing, and emphasizing safety constraints. It marks a significant leap forward in battery management technology, offering a safe, efficient, and cutting-edge solution to meet the growing demand for reliable energy storage

To provide a clear roadmap of this manuscript, Section \ref{sec:fundamentals} explains the fundamentals. In Section \ref{sec:dagger_app}, we describe the DAGGER-based approach for optimal constrained battery charging. Section \ref{sec:results} presents the simulation results in comparison to traditional MPC. Finally, Section \ref{sec:conclusions} outlines the findings derived from this paper.

\section{Fundamentals}\label{sec:fundamentals}

This section aims to provide a concise yet comprehensive exploration of essential concepts within the intricate realm of battery charging.

\subsection{Battery model}\label{sub:batt_model}

In lithium-ion battery modeling, the widely adopted single particle model (SPM)  \cite{Santhanagopalan2006} offers a balance between accuracy and computational efficiency. Derived from the Doyle-Fuller-Newman model, it treats electrodes as spherical particles and has proven accuracy in various studies \cite{Moura2017,zhang2022beyond}. Our model incorporates dual-state thermal dynamics from \cite{Perez2017a_spme_2_temperature}. For detailed equations, refer to \cite{Pozzi2022_ccta}.

Central to our analysis, a key variable is the  the battery's state of charge, $\text{SoC}(t)$, which varies over time according to
\begin{align}\label{eq:soc}
\frac{d\,\text{SoC}(t)}{d\,t} = \frac{I(t)}{3600 \cdot C_{batt}},
\end{align}
where $t$ is time, $I(t)$ indicates the current applied to the battery (with charging associated with positive current values), and $C_{batt}$ stands for the battery's capacity in $[\text{Ah}]$. The battery is considered fully charged when $\text{SoC}(t)=1$ (the maximum SoC value) and completely discharged when $\text{SoC}(t)=0$.

Certainly, SoC(t) within a battery plays a pivotal role in shaping the average stoichiometry levels within its electrodes, thereby impacting the overall performance of the battery. Surface stoichiometry, another critical aspect, is intricately linked to factors like concentration flux and applied current. The dynamic interplay of these factors is governed by coefficients derived from the battery's physical properties. It is noteworthy that these relationships can vary between different types of batteries, as discussed in \cite{Pozzi2022_ccta}.

The equation that characterizes the voltage across the battery, $V(t)$, is given by
\begin{align}\label{eq:voltage}
V(t) = U_p(t)-U_n(t)+\eta_p(t)-\eta_n(t)+R_{sei}I(t).
\end{align}
Within this expression, the open circuit potential and overpotential are represented by $U_i(t)$ and $\eta_i(t)$, for $i\in\{n,p\}$, the open circuit potentials are commonly expressed as non-linear functions that depend on the surface stoichiometry as detailed in \cite{Pozzi2022_ccta} and the overpontentials are defined as in \cite{Pozzi2020}. The voltage across the solid electrolyte interphase (SEI) resistance is described by the term $R_{sei}I(t)$. It is worth highlighting that the subscript $p$ is used to specify characteristics or conditions associated with the cathode, while the subscript $n$ is employed to denote those related to the anode of the battery.

The battery's thermal dynamics are represented by the two-state framework outlined in \cite{Perez2017a_spme_2_temperature}, which considers both the core and surface temperatures, denoted as $T_c(t)$ and $T_s(t)$, respectively. The thermal behavior is mathematically defined as follows:
\begin{subequations}\label{eq:temperature_dynamics}
\begin{align}
C_c\frac{d\,T_c(t)}{d\,t} & =\dot{Q}(t)-\frac{T_c(t)-T_s(t)}{R_{c,s}}, \\
C_s\frac{d\,T_s(t)}{d\,t} & = \frac{T_c(t)-T_s(t)}{R_{c,s}}-\frac{T_s(t)-T_{env}}{R_{s,e}},
\end{align}
\end{subequations}
in which $R_{c,s}$ and $R_{s,e}$ represent the thermal resistances from the core to the surface, and from the surface to the environment, respectively. Meanwhile, $C_c$ and $C_s$ are used to specify the heat capacities of the core and surface, $T_{env}$  describes the ambient temperature and $\dot{Q}(t)$ quantifies the heat generation, determined as
\begin{align}\label{eq:heat_generation}
\dot{Q}(t)= |I(t)(V(t)- U_p(t)+U_n(t))|.
\end{align}
Importantly, the nominal electrochemical parameters have been determined through experimental characterization of the Kokam SLPB 75106100 commercial cell, as extensively discussed in \cite{Ecker2015a}. The thermal parameters are drawn from the information provided in \cite{Perez2017a_spme_2_temperature}.

\subsection{Model Predictive Control for battery charging}\label{sub:tpc}

MPC addresses the optimization problem associated with fast battery charging, which involves maintaining efficient state of charge tracking while ensuring safety. Safety constraints, crucial in battery charging, prevent issues like overheating and damage. Mathematically, this scenario is framed as a constrained optimization problem to find the best charging protocol. Within this framework, we treat the battery as a discrete-time system under digital control, where piece-wise constant inputs are applied at specific discrete times $t_k,\, k \in \mathbb{N}$, sampled at intervals of $t_s$. The input current exerts influence on key state variables, including state of charge, temperature, and voltage. Precisely, at each time instance $t_k$, we determine the sequence of optimal currents $\mathbf{I}^\star_{[t_k,t_{k+H}]}$ for the next $H$ time steps $[t_{k},t_{k+1},\ldots,t_{k+H-1}]$. 

\vspace{4pt}
This constrained optimization problem is formally expressed as

\vspace{-0.5cm}

\begin{align}\label{eq:battery_problem}
\mathbf{I}^\star_{[t_k,\,t_{k+H}]} = \argmin_{\mathbf{I}_{[t_{k},t_{k+H}]}} \quad & q_{\text{soc}}\sum_{i=k+1}^{k+H} (\text{SoC}(t_i)-\text{SoC}_{\text{ref}})^2 \nonumber \\
&+r\sum_{i=k}^{k+H-1} I(t_i)^2
\end{align}
 that for $i=k,\,k+1,\,\cdots,\,k+H-1$ is subject to:
\begin{subequations}\label{eq:battery_limits}
\begin{alignat}{2}
& \text{battery dynamics in \eqref{eq:soc}--\eqref{eq:heat_generation}}\\
 &I_{\text{min}}\leq I(t_i)\leq I_{\text{max}}\\
 &\text{SoC}_{\text{min}}\leq \text{SoC}(t_i)\leq \text{SoC}_{\text{max}}\\
&T_c(t_i)\leq T_{\text{max}}\\
&T_s(t_i)\leq T_{\text{max}}\\
&V(t_i)\leq V_{\text{max}}
\end{alignat}
\end{subequations}
Here, the prediction horizon denoted as $H$ dictates the duration over which the control strategy is planned. Designers can tailor the control problem by adjusting coefficients like $q_{\text{soc}}$ and $r$, allowing them to prioritize objectives such as achieving the desired state rapidly, minimizing control effort, or striking a trade-off between the two. Additionally, intervals $[I_{\text{min}}, I_{\text{max}}]$ and $[\text{SoC}_{\text{min}}, \text{SoC}_{\text{max}}]$ define the feasible ranges for applied current and state of charge, respectively. Upper bounds for temperature ($T_{\text{max}}$) and voltage ($V_{\text{max}}$) are set to ensure safe operation.

For analysis purposes, we define a prediction horizon of $H=4$ and a time step of $t_s=10s$. Weight factors for the cost function are established as $q_{\text{soc}}=1$ and $r=10^{-6}$. Likewise, specific lower and upper bounds for the variables under study can be found in Table~\ref{tab:bounds}. Lastly, the reference of state of charge ($\text{SoC}_\text{ref}$) has not been specified, as it depends on individual charging preferences.
\begin{table}[bh]
    \centering
    \vspace{-0.25cm}
    \caption{The lower and upper limits of the primary battery variables}
    \vspace{-0.2cm}
    \begin{tabular}[width=0.1\textwidth]{ccc}
        \toprule[0.3mm]
        \textbf{Parameter} & \textbf{Minimum} & \textbf{Maximum}\\
        \midrule[0.3mm]
        Current&	-10 [A]&	10 [A]\\
        Voltage&	n.d&	4.2 [V]\\
        State of charge &	0&	1\\
        Temperature &	n.d&	313.15 [K]\\
         \bottomrule[0.3mm]
    \end{tabular}
    \label{tab:bounds}
    \vspace{-0.2cm}
\end{table}

In this context, MPC computes the optimal control sequence $\mathbf{I}^\star_{[t_{k},t_{k+H}]}$ over the prediction horizon $H$. This is achieved by solving the constrained optimization problem in \eqref{eq:battery_problem} - \eqref{eq:battery_limits}, the cost function of which is dependent on the predictions offered by a mathematical model of the battery. Subsequently, in alignment with the receding horizon paradigm, only the first element $I^\star(t_{k})$ of the resulting optimal input sequence is applied. At each time-step $t_k$, the MPC algorithm determines the optimal action $I^\star(t_k)$ based on the battery's model parameters $\mathbf{p}$ and the battery states vector $\mathbf{s}_{t_k}$. These parameters, $\mathbf{p}$, are specific to each battery cell and may change over time due to factors like aging. However, it is essential to note that in real-world scenarios, the controller typically lacks direct access to the internal states and electrochemical parameters, such as lithium-ion concentration, diffusion coefficient, electrolyte conductivity and so forth. In lieu of that, it can directly measure only voltage, surface temperature, and current.

By assuming perfect knowledge of states and parameters, an optimal charging sequence $\pi^\star$ can be obtained. This strategy maps the battery states and parameters to the optimal current, as follows:
\begin{align}
   \pi^\star :\,\, (\mathbf{s}_{t_k},\,\mathbf{p}) \rightarrow I^\star(t_k).
\end{align}
This optimal policy serves as the ``expert agent" in the imitation learning approach proposed in this manuscript.

\subsection{Imitation Learning}\label{sub:imit_learn}

Imitation learning, a machine learning approach, aims to replicate expert behavior without explicit knowledge of system dynamics. It learns from expert demonstrations and seeks to mimic the expert's behavior closely. It can be regarded as a type of supervised learning that establishes a policy, denoted $\pi_{\boldsymbol{\theta}}$, to accurately imitate an expert policy $\pi^\star$. This policy maps states from $\mathcal{S}$ to actions in $\mathcal{A}$, implemented through a machine learning model.

The learning process involves collecting pairs of states and actions $(s,a) \in \mathcal{S} \times \mathcal{A}$ through interactions with the environment or a suitable simulator.

From a perspective of supervised learning, its core objective is to establish a robust mapping  from states (inputs) to actions (targets) using a dataset $\mathcal{D} = \{(\mathbf{s}_{k}, \pi^\star(s_k))\}_{k=1}^{N}$ of state-action pairs generated by the expert policy. To align the imitation learning policy with the expert, a common loss function is the mean squared error between $\pi_{\boldsymbol{\theta}}$ and $\pi^\star$, expressed as:
\begin{align}\label{eq:loss}
L(\boldsymbol{\theta}) = \frac{1}{N} \sum_{k=1}^{N} \left[ \pi_{\boldsymbol{\theta}}(\mathbf{s}_{k}) - \pi^\star(\mathbf{s}_{k}) \right]^2.
\end{align}
Yet, behavioral cloning, a method used in training models, neglects the changing distribution of states, where the learned policy may encounter new situations deviating from the expert's actions. This oversight can lead to decreased performance and accumulating errors over time. Recognizing this challenge, the DAGGER method is introduced as a significant advancement, effectively addressing concerns related to distributional shift in the subsequent sections.

\section{DAGGER-based Imitation Learning Approach for Optimal Constrained Battery Charging}\label{sec:dagger_app}

This section first introduces the DAGGER algorithm and then discusses its application for constrained battery charging.

\subsection{Dataset Aggregation - DAGGER algorithm}
Dataset Aggregation, commonly referred to as DAGGER, is an algorithmic approach designed to address the issue of distributional shift in behavioral cloning \cite{Ross2011dagger}. In each iteration, DAGGER  systematically collects state-action pairs from two sources: firstly, the current learned policy, denoted as $\pi_{\theta_{i-1}}$, and secondly, the expert policy, expressed as $\pi^\star$. These pairs are then seamlessly incorporated into an augmented training dataset, laying the groundwork for subsequent policy updates. 

To efficiently gather data, DAGGER employs a mixed policy denoted as $\pi_i$, which combines actions from both the expert policy and the current learned policy. The decision to follow either policy is determined by the probabilities $\beta_i$ and $(1-\beta_i)$. These probabilities change at each iteration, dynamically adjusting the emphasis on either the expert or current learned policy based on the assigned likelihood values.
Mathematically, the mixed policy $\pi_i$ is defined as:
\begin{align}
\pi_i(\mathbf{s}_{t_k}) = 
\begin{cases}
\pi^\star(\mathbf{s}_{t_k}), & \text{with probability }\beta_i \\
\pi_{\theta_{i-1}} (\mathbf{s}_{t_k}), & \text{with probability }(1-\beta_i)\end{cases}
\end{align}
As a result, the contribution of the expert policy gradually diminishes over iterations, allowing the learned policy to make more autonomous decisions. 
After gathering data with the mixed policy, the learned policy is updated to $\pi_{\theta_{i}}$. This update minimizes the empirical loss based on the aggregated dataset up to the current iteration.

\begin{algorithm}[bh]
\caption{DAGGER Algorithm} \label{alg:dagger}
\begin{algorithmic}[1]
\REQUIRE dataset $\mathcal{D}_0$, iterations number $n_D$, decay factor $\beta_i$
\FOR{$i=1$ to $n_D$}
\STATE Train policy $\pi_{\theta_{i-1}}$ on $\mathcal{D}_{i-1}$
\STATE Generate new dataset $\mathcal{D}_i$ by following policy $\pi_{i}$
\STATE Aggregate the datasets: $\mathcal{D}_{i} = \mathcal{D}_{i} \cup \mathcal{D}_{i-1}$
\ENDFOR
\STATE Return final policy $\pi_{\theta_{n_D}}$
\end{algorithmic}
\end{algorithm}

The key steps of the DAGGER algorithm are highlighted in  Algorithm~\ref{alg:dagger}. Additionally, here are some crucial points to emphasize:
\begin{enumerate}[leftmargin=*]
    \item \textbf{Initialization:} We start by training an expert agent to generate the initial dataset ($\mathcal{D}_{0}$). This dataset serves as the foundation for subsequent training and data collection. Additionally, we define two essential parameters: the number of iterations ($n_D$) and the decay factor ($\beta_i$) which changes at each iteration. These parameters are the inputs of our algorithm.
    \item \textbf{Training the Learned Policy:} We train the learned policy ($\pi_{\theta_{i-1}}$) using the dataset $\mathcal{D}_{i-1}$. In this training process, the states serve as inputs, and the actions become our targets for the learned policy.
    \item \textbf{Data Collection with Mixed Policy:} To generate the new dataset, $\mathcal{D}_i$,  both the expert agent and the learned policy encounter novel and unfamiliar conditions due to the evolving process behavior (e.g. parameter variation due to battery aging). Resulting in the creation of an entirely distinct scenario that differs from the previous context. We then apply the mixed policy $\pi_i$ to determine the actions taken in response to these new, unfamiliar circumstances. At each time step of this new situation, we randomly select which policy to follow based on the probabilities set by $\beta_i$. Initially, $\beta_i$ may favor the expert policy to gather high-quality data, but as training progresses, $\beta_i$ may decrease, shifting more weight toward the learned policy as it becomes more reliable. Importantly, this choice is not deterministic; it is influenced by randomness, which ensures that our new generated dataset comes from both the expert agent and the learned policy, facilitating exploration while adapting over time.
    \item \textbf{Dataset Aggregation:} We then combine this new dataset $\mathcal{D}_i$ with the previous one, $\mathcal{D}_{i-1}$. This aggregation strengthens our dataset, incorporating fresh states and accounting for uncertainties.
    \item \textbf{Iterative Improvement:} The iterative process of the DAGGER algorithm involves revisiting step 2, utilizing an enhanced dataset $\mathcal{D}_i$ to refine the learned policy, and repeating until the specified iterations ($n_D$) are complete. An exit condition could be set based on achieving a minimal threshold in the reduction of training error or loss between two successive iterations.
\end{enumerate}

DAGGER refines learning iteratively, aligning states experienced by the learned policy with the expert's distribution. This continual adjustment improves the learned policy's ability to emulate the expert across various state spaces. Compared to standard behavioral cloning, DAGGER consistently enhances performance, effectively addressing distributional shift challenges, including parameter uncertainties and novel scenarios not present in the training dataset within the context of imitation learning.

\section{Simulation Results}\label{sec:results}
In this section, we provide a thorough overview of the experimental procedures conducted to validate the proposed methodology's effectiveness.
\subsection{Dataset Generation and Training Stage}
\begin{figure*}[th!]
\centering
\subfloat[]{\includegraphics[width=0.33\textwidth]{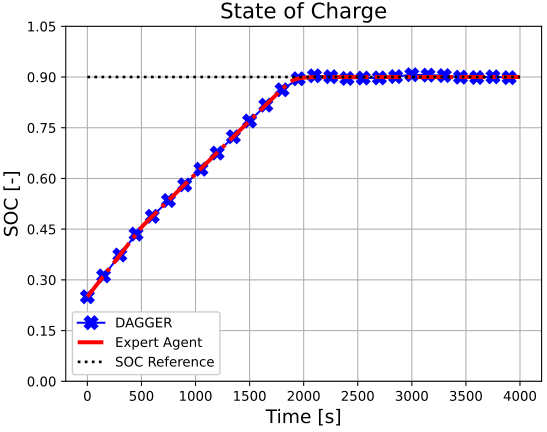}}\hfill
\subfloat[]{\includegraphics[width=0.33\textwidth]{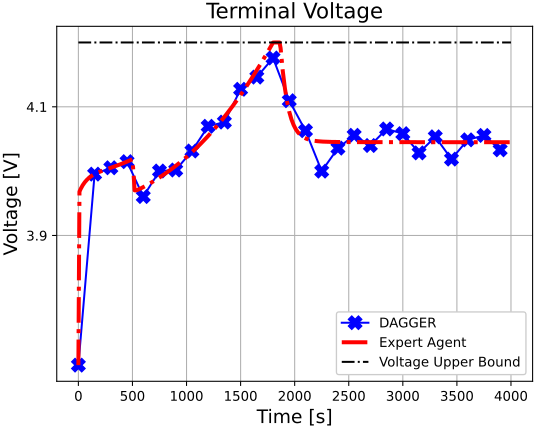}}\hfill
\subfloat[]{\includegraphics[width=0.33\textwidth]{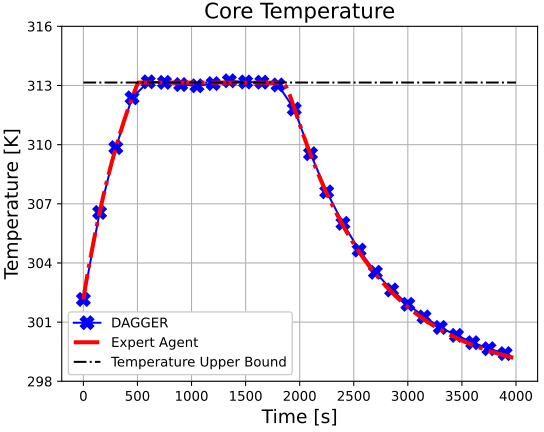}}
\hfill
\caption{Battery State Comparison: a) Battery state of charge comparison between the expert agent (dash-dotted red line) and the proposed DAGGER method (crossed blue line), showcasing satisfactory reference tracking by both control methods, b) Battery voltage profiles under the control of the expert agent (dash-dotted red line) and the proposed DAGGER method (crossed blue line) to ensure constraint satisfaction, c) Core temperature profiles to assess constraint compliance, with control strategies depicted in consistent colors and styles. }\label{fig:Battery_state_comparison}
\vspace{-0.4cm}
\end{figure*}
The initial dataset, $\mathcal{D}_0$, is generated through interactions between an expert MPC controller and a battery simulator. This dataset contains measurements such as voltage, temperature, applied current, expert-computed optimal current, and the reference state of charge for 500 episodes. Each episode spans 200 time-steps with a 10-second sample rate. To provide historical context, the dataset is structured to include measurements from the current time-step $t_k$ and past time-steps up to a window size of $n_W$.
The synthetic data allows for diversity by randomly sampling battery parameters for each episode, including state of charge ($\text{SoC}$), surface temperature ($T_s$), capacity ($C$), and SEI resistance ($R_{sei}$).
To improve realism, each episode starts with 30 steps where no current is applied, eliminating the need for initial control action estimates.
This dataset, $\mathcal{D}_0$, trains the initial policy $\pi_{\theta_0}$. For each iteration, policy $\pi_i$ is applied over 100 episodes, resulting in the dataset $\mathcal{D}_i$. This process is repeated for 15 iterations, leading to a dataset with 2000 episodes and the final policy $\pi_{\theta_{15}}$. The decay ratio $\beta_i$ (with $\beta_0=1$) decreases exponentially with each iteration following the guideline $\beta_i = 0.5\beta_{i-1}$.

The policy parameterization employs a Recurrent Neural Network (RNN), as described in \cite{Pozzi2023sensors}, due to its suitability for handling sequential data, a key characteristic of the dataset used in this study. This RNN model consists of four Long Short-Term Memory (LSTM) hidden layers (128, 64, 32, and 16 neurons) and four fully connected layers (100, 100, 50, and 10 neurons). To process the historical measurements, a window size ($n_W$) of 20 is adopted. The model utilizes a Rectified Linear Unit (ReLU) activation function in the hidden layers and a hyperbolic tangent (tanh) activation function in the output layer, which constrains the output within a specific interval to maintain optimal current within operational limits. A preprocessing pipeline involving feature scaling and standardization enhances model learning. Training employs stochastic gradient descent with the Adam optimizer and a mean squared error loss function, utilizing a learning rate of $5\times 10^{-4}$. To improve robustness and account for real-world sensor disturbances, Gaussian noise with standard deviations of 20 [mV] and 1 [K] for voltage and temperature, respectively, is added to training features.

This dataset generation and training approach ensures the DAGGER algorithm's adaptability to diverse real-world battery scenarios.

\subsection{Case of study of battery charging applying DAGGER-based method}

Here, we present a simulation designed to assess the proposed method's efficacy, with a primary emphasis on evaluating its benefits for tackling distributional shift in supervised learning challenges and its computational efficiency when compared to a conventional approach.

In the simulation, the objective is to charge a battery from an initial state of charge of 25\% to a target state of charge of 90\%. The battery's initial temperature is set at 302.5 [K] both at the core and on the surface. It is in a resting condition at the onset of the charging process. The battery is characterized by a solid electrolyte interface resistance of 0.0165 [$\Omega$] and a capacity of 6.75 [Ah]. The charging process should be completed within 4000 seconds while adhering to the constraints outlined in Table~\ref{tab:bounds}.
\begin{figure}[t!]
    \centering
    \includegraphics[width=0.75\columnwidth]{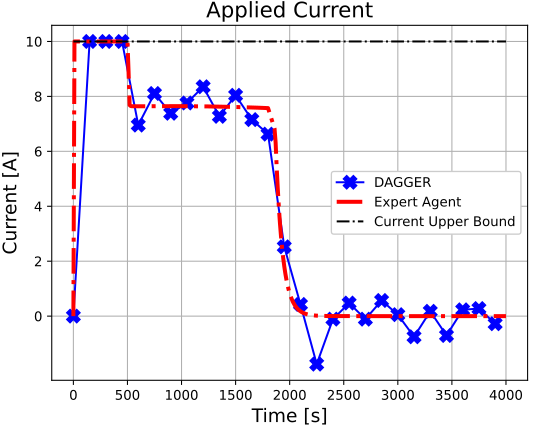}
    \vspace{-0.4cm}
    \caption{Comparison of Current Profiles: This figure presents a direct comparison of the current profiles generated by the expert agent (MPC) and the DAGGER approach over time. The expert agent's current profile is represented by a dash-dotted red line, while the profile of the DAGGER method is depicted as a crossed blue line. }
    \label{fig:current_dagger}
    \vspace{-0.6cm}
\end{figure}
To illustrate the performance differences between DAGGER and the expert agent, we present the evolution of key battery variables in Fig. \ref{fig:Battery_state_comparison}. DAGGER's performance is represented by the crossed blue line, while the expert agent's performance is depicted by the red dash-dotted line. The expert agent, an all-knowing MPC, optimizes battery management by enforcing voltage and temperature constraints. However, without prior battery knowledge, a standard MPC cannot ensure these constraints are met and might even exceed them. The proposed approach aims to mimic the omniscient MPC's performance using historical data, despite not guaranteeing constraint satisfaction. Nonetheless, it effectively meets the constraints in most instances.

Fig.~\ref{fig:current_dagger} depicts the dynamic current profile utilized by the DAGGER-based approach, which maintains an average value similar to that of the expert agent. The agent's fluctuating behavior, aimed at maintaining temperature constraints and charge states, is influenced by proactive control strategies for swift and safe battery charging, and potential distortions from Gaussian noise in historical data. This oscillation between conservative and aggressive actions may be mitigated by applying a low pass filter to voltage and temperature measurements, particularly effective when data is collected asynchronously at low sample rates, to reduce noise effects and enhance controller response accuracy.
\begin{figure}[t!]
    \centering
    \vspace{-0.1cm}
    \includegraphics[width=0.75\columnwidth]{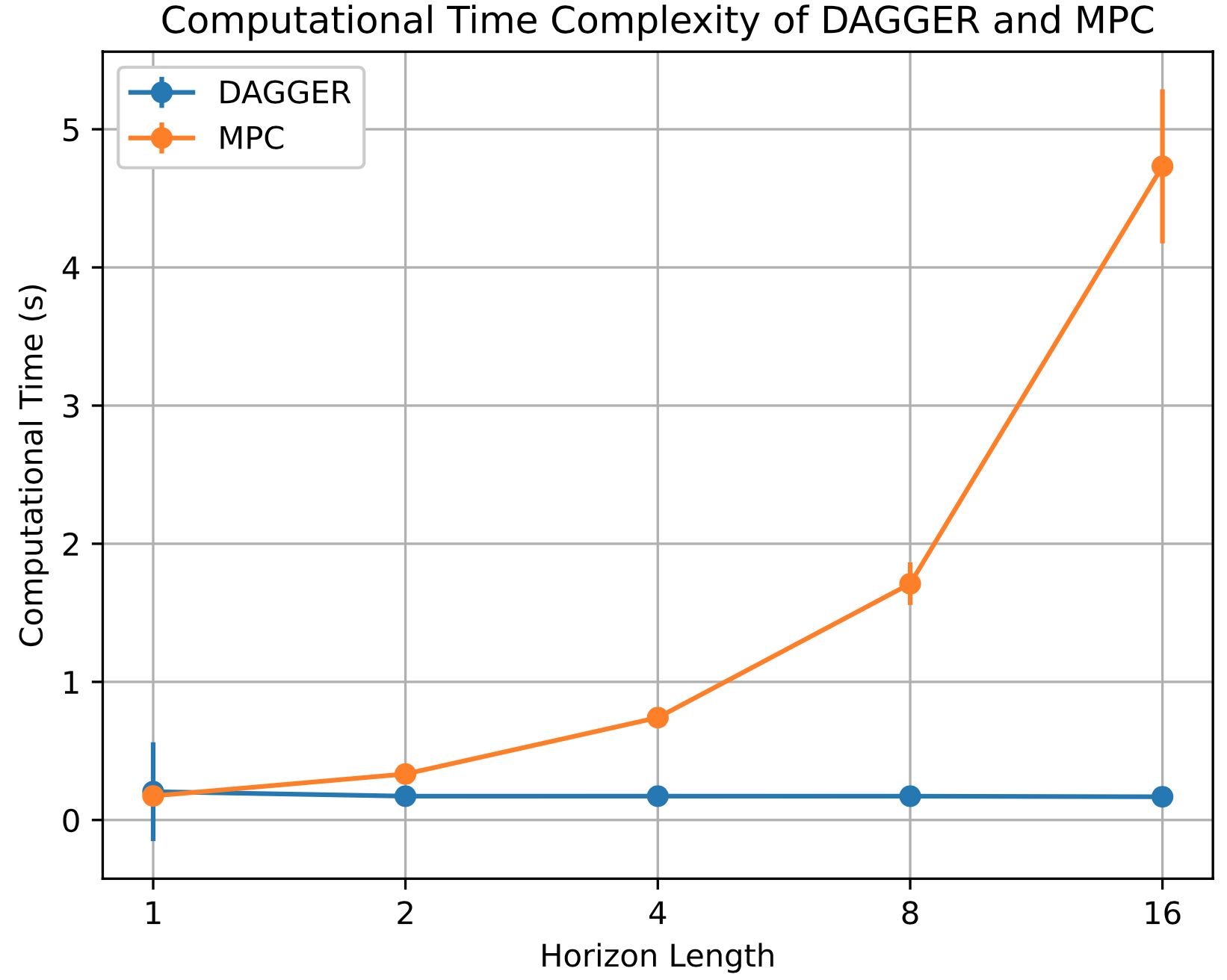}
    \vspace{-0.4cm}
    \caption{Online Computational Time Comparison: MPC (orange) vs. DAGGER (blue) with increasing prediction horizon.}
    \label{fig:comp_test}
    \vspace{-0.7cm}
\end{figure}
It is essential to note that the DAGGER algorithm exhibits comparable efficiency in handling voltage constraints. This is evident in Fig. 1(b), where the chattering effect of the DAGGER method's profile is chiefly due to high-frequency components present in the applied current, as discussed earlier. Notably, this oscillatory behavior does not have a substantial impact on the overall battery charging process

In conclusion, this study validates the effectiveness of DAGGER in the imitation learning paradigm for optimal constrained battery charging.

\subsection{Computational performance and statistical analysis}

In this subsection, we compare the online computational performance between the DAGGER-based algorithm and a traditional model predictive controller, used as a benchmark. We assess computational times for different prediction horizons $(H \in \{1, 2, 4, 8, 16\})$, conducting the test for both approaches at each horizon value. This analysis provides statistical insights into computational workloads.

Fig. \ref{fig:comp_test} presents the results, with solid lines corresponding to the average computational cost (blue for the DAGGER-based approach and orange for MPC). Remarkably, the standard predictive controller exhibits superlinear growth in computational time with rising prediction horizons, while the DAGGER's computational time remains fairly constant.

The proposed method stands out due to its minimal online efforts, requiring neural network evaluation in the measured state and a fixed horizon of 1 for recursive feasibility, in contrast to the standard predictive controller, which involves solving an optimization problem with expanding variables and constraints.

In 100 simulations comparing the DAGGER-based approach to the expert agent (MPC), we consider various battery parameters and initial states to demonstrate the proposed method's effectiveness in replicating expert agent performance across diverse battery conditions. The analysis involves comparing actions taken by DAGGER and the expert agent under identical battery states, highlighting that the DAGGER-based imitation learning operates with limited information compared to the expert agent's full knowledge.

\begin{figure}[th!]
    \centering
    \vspace{-0.25cm}
    \includegraphics[width=0.8\linewidth]{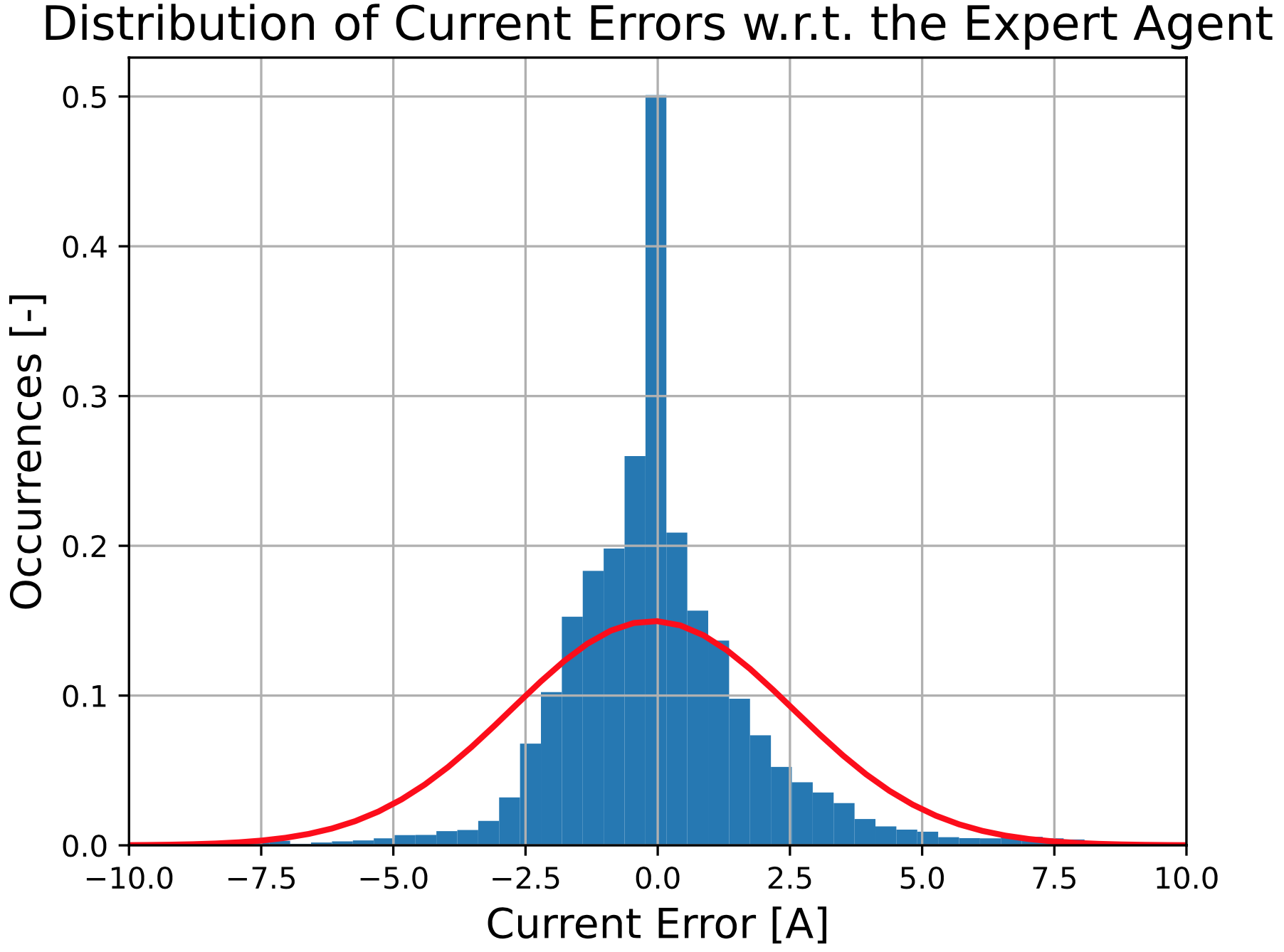}
    \vspace{-0.4cm}
    \caption{Distribution of current errors: A statistical analysis of DAGGER-based approach with respect to the expert agent (MPC).}
    \label{fig:errors-current}
    \vspace{-0.3cm}
\end{figure}

Fig.~\ref{fig:errors-current} depicts the distribution of current errors in emulating the optimal current applied by the expert agent. The results from the DAGGER framework exhibit a well-shaped Gaussian-like distribution (solid red line) with a mean close to zero (-0.03 [A]) and a standard deviation of 2.66 [A], indicating its exceptional capability to mimic the expert agent's actions across a broad spectrum of battery settings.

\section{Conclusions}\label{sec:conclusions}
In a rigorous simulation setting, the proposed DAGGER-based method has been compared to a traditional MPC approach. The results highlight its effective adaptability in emulating the expert agent across diverse battery conditions, and its enhanced computational performance. It excels in constraint handling and demonstrates robustness against uncertainties, making it a dependable and safer charging strategy.

Similarly, the DAGGER algorithm effectively tackles the distributional shift issue that is inherent in supervised learning and often results in safety constraint violations under varying conditions. Furthermore, it  maintains a balance between performance and safety across different scenarios. These findings underscore the potential of advanced imitation learning techniques, like DAGGER, in solving optimal control problems in situations with unmeasurable states and uncertain parameters.
\vspace{-0.1cm}
\bibliographystyle{IEEEtran}
\bibliography{dagger_conf}

\end{document}